\documentclass{amsart}
\usepackage{amsmath,latexsym}
\usepackage[psamsfonts]{amssymb}
\usepackage{times}
\usepackage[mathcal]{euscript}
\numberwithin{equation}{section}

\newcommand{\bbR}{\mathbb R}

\newcommand{\bbZ}{\mathbb Z}

\newcommand{\bbC}{\mathbb C}
\newcommand{\bbN}{\mathbb N}

\renewcommand{\epsilon}{\varepsilon}
\newcommand{\be}{\begin{equation}}
\newcommand{\ee}{\end{equation}}
\newcommand{\no}{\nonumber}

\newcommand{\spec}{\mathrm{spec}}
\newcommand{\cG}{{\mathcal G}}
\newcommand{\cH}{{\mathcal H}}

\newcommand{\Dom}{\mathrm{Dom}}
\newcommand{\dom}{\mathrm{Dom}}

{\bf}{\it}
\newtheorem{theorem}{Theorem}[section]

\newtheorem{remark}[theorem]{Remark}
\newtheorem{example}[theorem]{Example}

\begin{document}

\title{On $\mu$-scale invariant operators}

\author{K. A. Makarov}
\address{
Department of Mathematics, University of Missouri, Columbia, MO
65211, USA} \email{makarov@math.missouri.edu}

\author{E. Tsekanovskii }
\address{
 Department of Mathematics, Niagara University,
 NY 14109, USA} \email{tsekanov@niagara.edu}

\dedicatory{Dedicated to the memory of M. Krein on the occasion of
his one hundredth birthday anniversary}



\begin{abstract}

We introduce the concept of  a $\mu$-scale
invariant  operator with respect  to a unitary transformation in  a separable
 complex Hilbert space. We  show that if a nonnegative
densely defined symmetric operator is $\mu$-scale invariant for some $\mu>0$, then both the Friedrichs
and the Krein-von Neumann extensions of this operator  are also $\mu$-scale invariant.

\end{abstract}
  \maketitle

%%%%%%%%%%%%%%%%%%%%%%%%
\section{Introduction}

Given a unitary operator $U$ in a separable complex Hilbert space $\cH$ and a (complex) number
 $\mu\in \bbC\setminus \{0\}$, we introduce the concept of
   a $\mu$-scale invariant  operator $T$  (with respect to the transformation $U$) as a (bounded)  ``solution''
   of the following equation
\begin{equation}\label{bukv}
UTU^*=\mu T.
\end{equation}
Note,  that in this case
 $U$ and $T$ commute up to a  factor, that is,
 \begin{equation}\label{bukk}
UT=\mu TU,
\end{equation}
and then necessarily $|\mu|=1$ (see \cite{BBP}),  provided that
 $T$ is a bounded operator and
 $$\spec(UT)\ne \{0\}.
 $$

 The search for
 pairs of
unitaries $U$ and $T$ satisfying  the canonical (Heisenberg)
commutation relations \eqref{bukk}  with $|\mu|=1$ leads to
realizations of the rotation algebra, the $C^*$-algebra generated by
the monomials $T^mU^n$, $m,n\in \bbZ$ (see, e.g.,
 \cite{Sei}).
The irreducible representations of this algebra
 play a crucial  role in the  study   of the Hofstadter type models. For instance, the Hofstadter Hamiltonian
 $H=T+T^*+U+U^*$
  typically has  fractal spectrum
    that  is  rather sensitive  to the algebraic properties of the ``magnetic
  flux'' $\theta$,   $\mu=e^{i\theta}$,
   which is captured in the beauty of the famous Hofstadter butterfly (see \cite{Sei} and references therein).
  We also note that self-adjoint realizations $U$ and $T$  of commutation relations \eqref{bukv} or
  \eqref{bukk}  for $|\mu|=1$
  are obtained  in \cite{BBP} while the case of  contractive   (not necessarily self-adjoint) solutions
  $T$,  and unitary $U$, has been discussed  in \cite{PT}.

To incorporate the case of $|\mu|\ne 1$,
 where
 unbounded solutions to \eqref{bukv} are of necessity considered,  we extend
the concept of the  $\mu$-scale invariance
 to the case of unbounded operators $T$  by the requirement  that  $\Dom(T)$ is invariant, that is,
\begin{equation}\label{dom}
U^*\dom (T)\subseteq\dom (T),
\end{equation}
 and
\begin{equation}\label{self}
UTU^*f=\mu Tf \quad \text{ for all } f\in \Dom(T).
\end{equation}

In this short Note we restrict ourselves to the case  $\mu>0$ and
focus on the study   of symmetric as well as self-adjoint  unbounded
solutions $T$ of \eqref{dom} and \eqref{self}. Our main result (see
Theorem \ref{main})
 states that if a densely defined nonnegative
(symmetric) operator $ T$ is $\mu$-scale invariant with respect to
a unitary transformation $U$,  then the  two classical extremal
nonnegative self-adjoint extensions, the Friedrichs and the
Krein-von Neumann extensions,  are $\mu$-scale invariant as well.

 The paper is organized as follows: In Section 2, based on a result by Ando and Nishio
 \cite{AN}, we provide  the proof of Theorem \ref{main}. Section 3
 is devoted to further generalizations and a discussion of the $\mu$-scale
invariance concept from the standpoint of group representation
theory.

%%%%%%%%%%%%%%%%%%%%%%%%%%%%%%%%%%%%%%%%%%%%%%%%%%%%%%%%%%%%%%%%%%%%%%%%%%%%

%%%%%%%%%%%%%%%%%%%%%%%%%%%%%%%%%%%%%%%%%%%%%%%%%%%%%%%%%%%%%%%%%%%%%%%%%%%%
%%%%%%%%%%%%%%%%%%%%%%%%%%%%%%%%%%%%%%%%%%%%%%%%%%%%%%%%%%%%%%%%%%%%%%%%%%%%

%%%%%%%%%%%%%%%%%%%%%%%%%%%%%%%%%%%%%%%%%%%%%%%%%%%%%%%%%%%%%%%%%%%%%%%%%%%%

%%%%%%%%%%%%%%%%%%%%%%%%%%%%%%%%%%%%%%%%%%%%%%%%%%%%%%%%%%%%%%%%%%%%%%%%%%%%
%%%%%%%%%%%%%%%%%%%%%%%%%%%%%%%%%%%%%%%%%%%%%%%%%%%

%%%%%%%%%%%%%%%%%%%%%%%%%%%%%%%%%%%%%%%%%%%%%%%%%%%%%%%%%%%%%%%%%%%%%%%

\section{Main result}

 Recall that if $\dot A$ is a densely defined (closed) nonnegative operator, then the
 set of all nonnegative self-adjoint extensions of $\dot A$ has the minimal element  $A_K$,
 the Krein-von Neumann extension (different authors refer to  the minimal extension $A_K$
  by using different names,
 see,  e.g., \cite {AS}, \cite{AN}, \cite {AT}, \cite{Bir}),  and
 the  maximal one $A_F$, the Friedrichs extension. This means, in particular,  that  for any nonnegative self-adjoint extension
 $\tilde A$  of $\dot A$ the following operator inequality holds \cite{Kr1}
 $$
 (A_F+\lambda I)^{-1}\le (\tilde A+\lambda I)^{-1}\le (A_K+\lambda I)^{-1}, \quad
 \text{ for all } \lambda >0.
 $$

 The following  result characterizes the  Friedrichs and the
 Krein-von Neumann extensions a form convenient for our considerations.

 \begin{theorem}[\cite{AkG}, \cite{AN}]\label{ando}
  Let $\dot A$ be a (closed) densely defined nonnegative symmetric operator. Denote by $\bf a$
  the closure\footnote { Recall that  $f\in \Dom[{\bf a}]$ if and only if there exists a
  sequence $\{f_n\}_{n\in \bbN}$, $f_n\in \Dom(\dot A)$, such that\\$\lim_{n,m\to \infty}
  \dot {\bf a}[f_n-f_m]=0$ and $\lim_{n\to \infty} f_n= f.$}
 of the quadratic form
 \begin{equation}\label{kvf}
 {\dot {\bf a}}[f]=(\dot A
f, f), \quad \Dom[{\dot {\bf a}}]=\Dom(\dot A).
 \end{equation}
 Then,
\begin{itemize}
\item[ (i)]  the Friedrichs extension $A_F$ of $\dot A$  coincides with
 the restriction of the adjoint operator
 $\dot A^*$ on the domain
 $$
 \Dom (A_F)=\Dom (\dot A^*)\cap \Dom [{\bf a}];
 $$

 \item[(ii)] the Krein-von Neumann  extension $A_K$  of $\dot A$  coincides with
 the restriction of the adjoint operator
 $\dot A^*$ on the domain
  $
 \Dom (A_K)
 $
 which consists of the set of elements $f$ for which there exists a
 sequence $\{f_n\}_{n\in \bbN}$, $f_n\in \Dom(\dot A)$, such that
 $$
 \lim_{n,m\to \infty}{\bf a}[f_n-f_m]=0
 \quad \text{and}\quad
  \lim_{n\to \infty}\dot Af_n=\dot A^*f.
 $$
\end{itemize}

 \end{theorem}

We now state the main result of this Note:

\begin{theorem}\label{main}
Assume that $\mu>0$ and  that   a densely defined (closed)
nonnegative symmetric operator  $\dot A$ is $\mu$-scale invariant
with respect to a unitary transformation $U$; that is,
$$
U^*\Dom(\dot A)\subseteq \Dom(\dot A) )
$$
and that
$$
U\dot A U^*=\mu \dot A\quad \text{on } \dom(\dot A).
$$

Then
\begin{itemize}
\item[(i)] the adjoint operator $\dot A^*$,

 \item[(ii)] the
Friedrichs extension $A_F$ of $\dot A$, and

\item[(iii)] the  Krein-von Neumann extension $A_K$ of $\dot A$
\end{itemize}
are  $\mu$-scale invariant with respect to the unitary transformation $U$.
\end{theorem}
\begin{proof}

Clearly, it is sufficient to prove (i) followed by the proof of the
fact that the domains of both the Friedrichs and the
 Krein-von Neumann
extensions are invariant with respect to  the
operator $U^*$.

(i). Given $f\in \Dom (\dot A)$
 and $h\in \Dom(\dot A^*)$,  one obtains
\begin{align*}
(\dot Af, U^*h)&=(U\dot Af, h)=(U\dot AU^*Uf, h)
\no \\&=(\mu\dot AUf,h)=(Uf,\mu\dot A^*h)=(f,U^*\mu \dot A^*h),
\end{align*}
thereby proving the inclusion $U^*\dom (\dot A^*) \subseteq \dom (
\dot A^*)$ as well as the equality
\begin{equation}\label{adj}
\dot A^*U^*h=\mu U^*\dot A^*h,\quad   h\in\dom(\dot A).
\end{equation}
The proof of (i) is complete.

(ii). First we show that
the domain of the closure of the quadratic
 form \eqref{kvf} is invariant with respect to operator the
  $U^*$.

 Recall that  $f\in \Dom[{\bf a}]$ if and only if there exists a
  sequence $\{f_n\}_{n\in \bbN}$, $f_n\in \Dom(\dot A)$, such that
 $$
 \lim_{n,m\to \infty} \dot{\bf a}[f_n-f_m]=0 \quad \text{and}\quad
 \lim_{n\to \infty} f_n= f.
 $$
  Take an $f\in \Dom[{\bf a}]$ and a sequence $\{f_n\}_{n\in \bbN}$
  satisfying the properties above.
 Clearly
 \begin{equation}\label{uf}
 \lim_{n\to \infty} U^*f_n= U^*f,
 \end{equation}
 with $U^*f_n\in \Dom(\dot A)$.
  Moreover,
 \begin{align*}
 {\bf a}[U^*f_n-U^*f_m]&=(\dot AU^*(f_n-f_m),U^*(f_n-f_m))
=
 (\dot UAU^*(f_n-f_m),(f_n-f_m))
 \\ &=\mu(\dot A(f_n-f_m),(f_n-f_m))
 = \mu{\bf a}[f_n-f_m].
 \end{align*}
Since $
 \lim_{n,m\to \infty}{\bf a}[f_n-f_m]=0
 $,
  one proves that
 $$
 \lim_{n,m\to \infty}{\bf a}[U^*f_n-U^*f_m]=0
 $$
 which together with \eqref{uf} implies that
  $U^*f\in \Dom[{\bf a}]$.
 Hence, we have proven the inclusion
 \begin{equation}\label{incl}
 U^*\Dom[{\bf a}]\subseteq \Dom[{\bf a}].
 \end{equation}

 Next,
  by (i) the domain $\dom(\dot A^*)$ is invariant with respect to $U^*$. This combined with \eqref{incl}
and Theorem \ref{ando} (i) proves that the domain of the Friedrichs
extension $A_F$ of $\dot A$ is invariant with respect to
the
 operator
$U^*$. Therefore, $A_F$ is $\mu$-scale invariant as a restriction of
the $\mu$-scale invariant operator $\dot A^*$ onto a $U^*$-invariant  domain.

(iii). Analogously, in order to show that the  Krein-von Neumann
extension $A_K$ is $\mu$-scale invariant with respect to the
transformation $U$,
 it is sufficient to show  that its domain is invariant with respect to $U^*$.

 Take $f\in \Dom (A_K)$. By Theorem \ref{ando} (ii)  there exists an
  ${\bf a}$-Cauchy sequence\footnote{ in the ``metric'' generated by the form ${\bf a}$
}
 $\{f_n\}_{n\in \bbN}$,
 $f_n\in \Dom(\dot A)$,
 such that
 \begin{equation}\label{cauchy}
 \lim_{n\to \infty}\dot A f_n=\dot A^*f.
 \end{equation}
From \eqref{adj} it follows that
\begin{equation}\label{figi}
\dot A U^*f_n=\dot A^*U^*f_n=\mu U^*\dot A^* f_n=\mu U^*\dot Af_n\quad \text{and}
\quad \dot A^*U^*f=\mu
 U^*\dot A^*f.
\end{equation}
Combining \eqref{cauchy} and \eqref{figi}, for
 the  ${\bf a}$-Cauchy sequence  $\{U^*f_n\}_{n\in \bbN}$ one gets
  \begin{equation*}
 \lim_{n\to \infty}\dot A U^*f_n =
\mu
 U^*\dot A^*f
=\dot A^*U^*f
 \end{equation*}
proving  that  $U^*f\in \Dom(A_K)$ by Theorem \ref{ando} (ii).Thus,
  $\Dom(A_K)$ is $U^*$- invariant and,
  therefore, the Krein-von Neumann extension $A_K$ is $\mu$-scale invariant as a restriction of the
  $\mu$-scale invariant operator $\dot A^*$  onto a $U^*$-invariant  domain.
 \end{proof}
\begin{remark} We remark that the concept of $\mu$-scale invariant operators can immediately be extended to the case of liner relations: we say that a linear relation $S$ is $\mu$-scale invariant with respect to the unitary transformation $U$ if its domain is $U^*$-invariant and $(f,g)\in S$ implies $(U^*f,\mu U^*g)\in S$.

Recall that the Friedrichs extension $S_F$ of a semi-bounded from
below  relation $S$ is defined as the restriction of $S^*$ onto the
domain of the closure of the quadratic form associated with the
operator part of $S$ \cite{cod} and the Krein-von Neumann extension
$S_K$ is defined by
 \begin{equation}\label{lrlr}
 S_K=\left(( S^{-1})_F  \right)^{-1},\end{equation}
provided that $S$ is, in addition,  nonnegative \cite{CS}  (no care should be taken about inverses, for they always exist).

Assume that a nonnegative  linear relation $S$ is $\mu$-scale invariant.
  Almost literally repeating the arguments of the proof of Theorem \ref{main} (i) one concludes that the adjoint relation $S^*$ is also $\mu$-scale invariant.
    Given the above characterization of the Friedrichs extension of a semi-bounded relation,  applying  Theorem \ref{main} (ii)  proves  the $\mu$-scale invariance
  of  $S_F$. As it follows from \eqref{lrlr}, a simple observation that
   $S$ is $\mu$-scale invariant if and only if the inverse relation  $S^{-1}$ is
 $\mu^{-1}$-scale invariant ensures that
 the Krein-von Neumann extension $S_K$ of $S$ is also $\mu$-scale invariant.
Thus, Part (iii) of Theorem \ref{main} is a direct consequence of Parts (i) and (ii) up to the  representation theorem  that states that Krein-von Neumann extension
$A_K$ of a nonnegative densely defined symmetric operator $\dot A$ can be ``evaluated" as
 \begin{equation}\label{krkr}
 A_K=\left(( \dot A^{-1})_F  \right)^{-1},\end{equation}
  with
  $\dot A^{-1}$  being  understood as a linear relation (for the proof of \eqref{krkr} we refer to \cite{CS}, also see
   \cite{AN} and
\cite{AT}).
\end{remark}
\begin{remark}\label{vottak}
Note without proof that if the symmetric nonnegative operator $\dot A$ referred to in Theorem \ref{main} has deficiency indices $(1,1)$ the Friedrichs and the Krein-von Neumann extensions of $\dot A$ are the  only ones $\mu$-scale invariant self-adjoint extensions.
\end{remark}
  The following simple example illustrates the statement of
  Theorem \ref{main}.

 \begin{example}  Assume that $\mu>0$, $\mu\ne1$, and that $U$ is the unitary scaling transformation
 on the Hilbert space $\cH=L^2(0,\infty)$ defined by
  $$ (Uf)(x)= \mu^{-\frac{1}{4}}f(\mu^{-\frac{1}{2}}x), \quad f\in L^2(0,\infty).$$
$T$ is the maximal operator on the Sobolev space $H^{2,2}(0,\infty)$
defined by
$$
T=-\frac{d^2}{dx^2},\qquad \Dom(T)=H^{2,2}(0,\infty).
$$
Let $A_F$ and $A_K$ be the restrictions of $T$ onto the domains
$$
\Dom(A_F)=\{f\in \Dom (T)\,|\,f(0)=0\}
$$
and
$$
\Dom(A_K)=\{f\in \Dom (T)\,|\,f'(0)=0\}
$$
 respectively. Denote by $\dot A$ the restriction of $T$ onto the domain
 $$
 \Dom(\dot A)=\Dom(A_F)\cap \Dom(A_K).
 $$
It is well known that $\dot A$ is a closed nonnegative symmetric
operator with deficiency indices $(1,1)$ and that $A_F$ and $A_K$
are the Friedrichs and the Krein-von Neumann extensions of $\dot A $
respectively and $T=\dot A^*$. A straightforward computation shows
that all  the operators $\dot A$, $A_F$,  $A_K$  and $T$ are
$\mu$-scale invariant with respect to the transformation $ U$.
Moreover, note that any other nonnegative self-adjoint extensions of
 $\dot A$ different from the extremal ones, $A_F$ and $A_K$, can be
obtained by the restriction of $T$ onto the domain (see, e.g., \cite{Nai},  also
see   \cite{DM} and
 \cite{DMF})
$$
\Dom(\tilde A_s)=\{f\in \Dom (T)\,|\,f'(0)=sf(0)\}, \quad\text{ for some } s>0,
$$
which is obviously not $U^*$-invariant.  Thus,
 the operator $\dot A$
 admits the
only  two $\mu$-scale invariant extensions,  the
 Friedrichs and the Krein-von Neumann extensions (cf. Remark \ref{vottak}).
 \end{example}

\section{  Concluding remarks}

 We remark that any  $\mu$-scale invariant  operator $T$ with respect to a  unitary transformation $U$
 is also $\mu^n$-scale invariant with respect to the (unitary) transformations  $U^n$,  $n=0, 1, ...\, $.
  That is,
 \begin{equation}\label{utut}
U^nTU^{-n}=\mu^n T,\quad \text{ for all } n\in \{0\}\cup\bbN.
\end{equation}
 If, in addition,
$$
U^*\dom (T)=\dom (T),
$$
then relation \eqref{utut} holds for all $n\in \bbZ$. Thus, we naturally arrive at a  slightly more general concept
of scale invariance
 with respect to a one-parameter  unitary representation
of the additive group $\cG$ ($\cG=\bbN$ or $\cG=\bbR$):
{\it Given a character $ \mu $,  $\mu:G\to \bbC$, of the group $\cG$
and its  one-parameter
 unitary representation  $g\mapsto U_g$,
a densely defined operator $T$ is said to be $\mu$-character-scale
 invariant with respect to the representation  $U_g$ if
$$
U_g\dom (T)=\dom (T),\quad g\in \cG,
$$
and
\begin{equation}\label{ututu}
U_gTU_{-g}=\mu(g) T,\quad \text{on} \quad \Dom(T), \quad g\in \cG.
\end{equation}
}

Clearly,  an appropriate  version
 of Theorem \ref{main} can almost literally  be restated in this
more general setting. It is also worth mentioning that upon
introducing the representation $V_g=\mu^gU_g$, $g\in \cG$,
 one can rewrite \eqref{ututu}
in the form
\begin{equation}\label{mumu}
U_gT=TV_g, \quad g\in \cG,
\end{equation}
and we refer  the interested reader to the papers
 \cite {LJ} and \cite{PT} where  commutation relations \eqref{mumu} for general groups $\cG$ with not
  necessarily unitary representations $U_g$ and $V_g$, $g\in \cG$, of the group $\cG$ are discussed.

Note that an infinitesimal analog  of the commutation relation in
\eqref{ututu} is also available provided that $\cG=\bbR$ and the
unitary representation $U_t$, $t\in \bbR$,
 is  strongly continuous.
In this case
 infinitesimal version of  \eqref{ututu}
 can heuristically be  written down  as the following commutation relation
\begin{equation}\label{lie}
[B,T]=i\hbar T,
\end{equation}
with  $[\cdot,\cdot]$  the usual commutator and
\begin{equation}\label{con}
\hbar=-\log \mu,
\end{equation}
the structure constant of the simplest noncommutative two-dimesional
Lie algebra \eqref{lie} and \eqref{con}. Here  $B$  is the
infinitesimal generator of the group $U_t$, so that $U_t=e^{iBt}$,
$t\in \bbR$. And in conclusion, note that Theorem \ref{main} paves
the way for  realizations of the Lie algebra
  by self-adjoint operators, provided that some ``trial'' symmetric
 realizations of the Lie algebra
are available.
\vskip 0.3cm
\noindent{\bf Acknowledgments}. We would like to thank   Steve Clark and Fritz Gesztesy for useful discussions.

\end{document}